\documentclass[prd,aps,letterpaper,floatfix,superscriptaddress,preprintnumbers,11pt,nofootinbib]{revtex4-1}
\usepackage{epsfig,epsf}
\usepackage{booktabs}
\usepackage{mathrsfs}
\usepackage{graphicx}
\usepackage{subcaption}
\usepackage{caption}
\usepackage{latexsym}
\usepackage{amsmath}
\usepackage{amssymb}
\usepackage{textcomp}
\usepackage{amsbsy}
\usepackage{tabularx}
\usepackage{slashed}
\usepackage{array}
\usepackage[dvipsnames]{xcolor}
\usepackage{bm}
\usepackage{amsmath,amsfonts,amssymb}
\usepackage{bbold}
\usepackage{hyperref}
\usepackage{float}
\usepackage{filecontents}
\DeclareGraphicsExtensions{.png,.eps,.pdf}

\usepackage{color}
\usepackage[normalem]{ulem}

\definecolor{darkgreen}{cmyk}{1,0,1,0.4}

\newcommand{\arXivold}[2]{\href{http://arxiv.org/pdf/#1}{{\tt #2/#1}}}
\newcommand{\arXiv}[2]{\href{http://arxiv.org/pdf/hep-ph/#1}{{\tt #2/#1}}}

\long\def\/*#1*/{}

\begin{document}

\title{Can the Higgs Still Account for the $g-2$ Anomaly?}
\author{Fayez Abu-Ajamieh}
\email{fayezajamieh@iisc.ac.in}
\affiliation{Centre for High Energy Physics, Indian Institute of Science, Bangalore 560012, India}
\author{Sudhir K. Vempati}
\email{vempati@iisc.ac.in}
\affiliation{Centre for High Energy Physics, Indian Institute of Science, Bangalore 560012, India}
\begin{abstract}
We use an Effective Field Theory (EFT) approach to evaluate the viability of the Higgs to account for the $g-2$ anomaly. Although the SM contribution of the Higgs to the muon's magnetic dipole moment is negligible, using a \textit{bottom-up} EFT, we show that given the current level of experimental limits on the Higgs sector, the Higgs can still yield a viable solution to the $g-2$ anomaly if its couplings to the rest of the SM particles are allowed to deviate from their SM predictions. Such a solution would only require an $O(1)$ fine-tuning. Further, applying unitarity arguments, we show that such a solution would indicate a scale of New Physics (NP) of $\sim 5 -8$ TeV, which could be lowered to $\sim 3.4 -4$ TeV if the Higgs couplings to the $W$ and $Z$ are assumed to conform to their SM predictions. We show that such a scenario could yield significant enhancement to the di-Higgs production in muon colliders, thus providing further motivation for its consideration. A key takeaway of this study is to highlight the importance of measuring the $\bar{\mu} \mu hh$ coupling in future experiments.
\end{abstract}
\maketitle

\section{Introduction}\label{Sec:intro}
The anomalous magnetic dipole moment of the muon, known as the $g-2$ anomaly, remains one of the best hints for physics Beyond the Standard Model (BSM). The E821 experiment at BNL \cite{Muong-2:2006rrc} first showed a discrepancy compared with the theoretical predictions, at a significance of $\sim 3.7 \sigma$. This was confirmed in 2021 by the E989 experiment at Fermilab \cite{Muong-2:2021ojo, Muong-2:2021ovs, Muong-2:2021vma}, which yielded a measurement of $a_{\mu}^{\text{EXP}} = 116592061(41)\times10^{-11}$. The Fermilab measurement, in comparison with the latest theoretical SM prediction of $a_{\mu}^{\text{SM}} = 116591810(43)\times10^{-11}$ \cite{Aoyama:2020ynm,Davier:2017zfy,Keshavarzi:2018mgv,Colangelo:2018mtw, Hoferichter:2019mqg,Davier:2019can,Keshavarzi:2019abf,Kurz:2014wya,FermilabLattice:2017wgj,Budapest-Marseille-Wuppertal:2017okr,RBC:2018dos,Giusti:2019xct,Shintani:2019wai,FermilabLattice:2019ugu,Gerardin:2019rua,Aubin:2019usy,Giusti:2019hkz,Melnikov:2003xd,Masjuan:2017tvw,Colangelo:2017fiz,Hoferichter:2018kwz,Gerardin:2019vio,Bijnens:2019ghy,Colangelo:2019uex,Pauk:2014rta,Danilkin:2016hnh,Jegerlehner:2017gek,Knecht:2018sci,Eichmann:2019bqf,Roig:2019reh,Colangelo:2014qya,Blum:2019ugy,Aoyama:2012wk,Aoyama:2019ryr,Czarnecki:2002nt,Gnendiger:2013pva}, further increases the discrepancy to $4.2\sigma$
\begin{equation}\label{eq:DeltaaExp}
\Delta a_{\mu} = a_{\mu}^{\text{EXP}} - a_{\mu}^{\text{SM}} = 251(59) \times 10^{-11}.
\end{equation}

On the other hand, some of the latest high-precision QCD lattice simulations \cite{Borsanyi:2020mff, Ce:2022kxy, Alexandrou:2022amy} appear to agree with the experimental measurements, thus reducing the anomaly. However, these results could be in conflict with the $e^{+}e^{-}$ data \cite{Colangelo:2022vok, DiLuzio:2021uty}. This discrepancy arises mainly from the Hadronic Vacuum Polaization (HVP) corrections, which are difficult to calculate (see for instance \cite{Crivellin:2020zul, Keshavarzi:2020bfy, Colangelo:2020lcg}). Therefore, it seems that further confirmation is needed. On the other hand, a new measurement of the cross section of $e^{+}e^{-} \rightarrow \pi^{+}\pi^{-}$ from the CMD-3 group \cite{CMD-3:2023alj}, reveals a larger contribution to $a_{\mu}^{\text{had,LO}}$ compared to the previous values in the literature, which makes $a_{\mu}^{\text{SM}}$ closer to the experimental value, thereby lowering the significance of the discrepancy in the data-driven approach as well. Nonetheless, this result too is inconsistent with the previous results in the literature and still needs independent corroboration.

The upcoming runs of the E989 experiment are expected to lower the uncertainty by a factor of 4, whereas the E34 collaboration at J-PARC \cite{Saito:2012zz, Mibe:2011zz,Abe:2019thb} plans on measuring $a_{\mu}$ using a completely different method. As for the SM predictions, various theory and lattice groups are expected to present updated results from those in \cite{Borsanyi:2020mff, Ce:2022kxy,Alexandrou:2022amy}. For the purposes of this study, we will assume that the anomaly exists and is given by eq. (\ref{eq:DeltaaExp}).

We should also point out that another discrepancy between the predicted and measured magnetic dipole moments of the electron was reported, however, there is disagreement between the two experiments that reported the discrepancy on the sign and significance of the electron $g-2$ anomaly. The first group \cite{Aoyama:2019ryr} reported an electron anomaly $\Delta a_{e} = -88(36) \times 10^{-14}$ corresponding to a significance of $2.4\sigma$, whereas the second group \cite{Morel:2020dww} reported $\Delta a_{e} = 48(30) \times 10^{-14}$, corresponding to a significance of only $1.6\sigma$. Given this disagreement, it is not clear whether the electron anomaly indeed exists or not, and if it does which sign it has. As the sign of the discrepancy is crucial in determining the nature of the physics behind the anomaly, we will ignore it in this work, waiting for more accurate measurements to be made, and for an agreement to be reached.

In this paper, we seek to investigate whether or not the Higgs boson can still accommodate a solution to the $g-2$ anomaly for the muon. In the SM, the contribution of the Higgs to $a_{\mu}$ at one loop is $O(10^{-14})$ and is thus negligible, however, the current level of experimental bounds on the Higgs couplings to the rest of the SM particles still leaves ample room for significant deviation from the SM predictions. For instance, the coupling of the Higgs to $\bar{\mu}\mu$ is only constrained at $O(1)$, whereas higher-dimensional Higgs operators remain essentially unconstrained. This suggests using an EFT approach in the Higgs interaction with the muon, in order to determine whether or not the muon anomaly can still be solved by the Higgs boson within the experimental constraints. The strategy is simple: We write down the most general Higgs interactions with the muon, and then calculate its contribution to $a_{\mu}$ given the experimental constraints, then we determine whether this contribution can account for the measured discrepancy in eq. (\ref{eq:DeltaaExp}) or not. Similar work using the Standard Model EFT (SMEFT) already exists in the literature \cite{Buttazzo:2020ibd,Yin:2020afe,Fajfer:2021cxa, Aebischer:2021uvt, Allwicher:2021jkr,Cheung:2021iev}, however, here we utilize a more phenomenologically-transparent, completely \textit{model-independent} approach \cite{Chang:2019vez, Abu-Ajamieh:2020yqi, Abu-Ajamieh:2021egq, Abu-Ajamieh:2022ppp, Abu-Ajamieh:2021vnh} in order to calculate the Higgs BSM contributions to the anomaly. We find that given the current level of experimental constraints, the Higgs sector can still provide a solution to the muon $g-2$ anomaly. We show that the viable part of the parameter space can be significantly probed by future experiments, especially the proposed $10$-TeV muon collider \cite{MuonCollider:2022cre}.

This paper is organized as follows: In Sec. \ref{Sec:Overview}, we present a review of the model-independent EFT used in \cite{Chang:2019vez, Abu-Ajamieh:2020yqi, Abu-Ajamieh:2021egq,Abu-Ajamieh:2022ppp,Abu-Ajamieh:2021vnh, Abu-Ajamieh:2023qvh}. In Sec. \ref{Sec:Contributions}, we present the contributions to $\Delta a_{\mu}$ in this framework at 1 and 2 loops, and we present the current experimental bounds and future projections on the parameter space. In Sec. \ref{Sec:Pheno} we discuss the phenomenological implications of the model. In Sect. \ref{sec:UVmodels}, we introduce a simple UV extension that can be matched to the EFT, and finally we conclude in Sec. \ref{Sec:Conc}.
\section{Overview of the Model-independent EFT}\label{Sec:Overview}
In this paper, we adopt the EFT framework employed in \cite{Chang:2019vez, Abu-Ajamieh:2020yqi, Abu-Ajamieh:2021egq,Abu-Ajamieh:2022ppp, Abu-Ajamieh:2021vnh, Abu-Ajamieh:2023qvh}. In this framework, we avoid any power expansion in writing down higher-dimensional operators, as is the case in the SMEFT. Instead, we parameterize NP as deviations from the SM predictions. Therefore, we write the effective Lagrangian of the Yukawa interaction of the lepton sector as follows\footnote{This methodology of expanding the Higgs couplings, is reminiscent of the chiral expansion of the Higgs Lagrangian (see for example,\cite{Buchalla:2020kdh} and the references therein).}

\begin{equation}\label{eq:BSM_Lag1}
\mathcal{L}_{\text{eff}} = \mathcal{L}_{\text{SM}} -\frac{Y_{l}v}{\sqrt{2}}\Big( \overline{L}_{l}\Tilde{\hat{H}} l_{R} + h.c.\Big)\Big[ \delta_{l1} \frac{X}{v} + C_{l2}\frac{X^{2}}{2!v^{2}}+ \dots \Big] + \ldots, 
\end{equation}
where $\delta_{l1}$ parameterize the deviations of the leptons' Yukawa couplings compared to the SM predictions
\begin{equation}\label{eq:Deviation}
\delta_{l1} \equiv \frac{Y_{l}-Y_{l}^{\text{SM}}}{Y_{l}^{\text{SM}}},\footnote{Notice that in the $\kappa$ framework, $\delta_{x} = \kappa_{x}-1$.}
\end{equation}
whereas $C_{ln}$ are Wilson coefficients that do not have SM counterparts. The field $X$ is defined in terms of the Higgs doublet $H$ as
\begin{equation}\label{eq:Xfield}
X \equiv \sqrt{2H^{\dagger}H} - v,
\end{equation}
whereas we define the projector $\Tilde{\hat{H}} \equiv \epsilon \hat{H}^{*}$, with
\begin{equation}\label{eq:Projector}
\epsilon = \begin{pmatrix} 
            0 && 1 \\ 
            -1 && 0
            \end{pmatrix}, \hspace{10mm}
            \hat{H} \equiv \frac{H}{\sqrt{H^{\dagger}H}} = \begin{pmatrix} 0 \\ 
                                                                    1 
                                                    \end{pmatrix} + O(\vec{G}),
\end{equation}
where $\vec{G}$ are the Goldstone bosons. Notice that we are dividing $X$ by the appropriate powers of $v$ only to keep the Wilson coefficients dimensionless, i.e., $v$ is NOT an expansion parameter. 
Notice that in the unitary gauge, $X \rightarrow h$ and the Goldstone bosons are eaten. In this paper, we simplify our calculations by working in the unitary gauge. Therefore, in this gauge, the Lagrangian in terms of the physical masses $m_{l} = \frac{Y_{l}v}{\sqrt{2}}$ becomes
\begin{equation}\label{eq:BSM_Lag2}
\mathcal{L}_{\text{eff}} = -m_{l}\overline{l}l \Big[ (1+\delta_{l1}) \frac{h}{v} + C_{l2} \frac{h^{2}}{2v^{2}} + \dots \Big],
\end{equation}
and for the case of the muon $l = \mu$. Before we proceed, a couple of remarks are in order. First, we are assuming that $v$ is the minimum of the Higgs potential including all higher-order corrections, i.e., $v=246$ GeV, and second, although eq. (\ref{eq:BSM_Lag2}) appears similar to the Higgs EFT (HEFT) \cite{Grinstein:2007iv}, we should keep in mind that secretly we are using the Higgs doublet in our expansion, and one can easily demonstrate that the effective Lagrangian in eqs. (\ref{eq:BSM_Lag1}) and (\ref{eq:BSM_Lag2}) can be mapped to any expansion in either the SMEFT or the HEFT (see \cite{Abu-Ajamieh:2020yqi, Abu-Ajamieh:2021egq,Abu-Ajamieh:2022ppp, Abu-Ajamieh:2021vnh} for the matching to the SMEFT operators). In other words, the deviations and Wilson coefficients in eqs. (\ref{eq:BSM_Lag1}) and (\ref{eq:BSM_Lag2}) can receive corrections from any number of higher-order operators in either the SMEFT or the HEFT frameworks. The benefit of this construction is twofold. First: there are fewer assumptions in this framework compared to either the SMEFT or the HEFT. Namely, we are only assuming that there are no light degrees of freedom below the energy scale where the EFT breaks down, and that the deviations and Wilson coefficients are compatible with experimental measurements. The second benefit lies in the fact that parameterizing NP this way is more phenomenologically-transparent, as these deviations and Wilson coefficients are what is measured experimentally as opposed to any expansion scale. 

\section{The BSM contributions to $(g-2)_{\mu}$}\label{Sec:Contributions}
\begin{figure}
\centering
\begin{subfigure}{0.15\textwidth}
    \includegraphics[width=\textwidth]{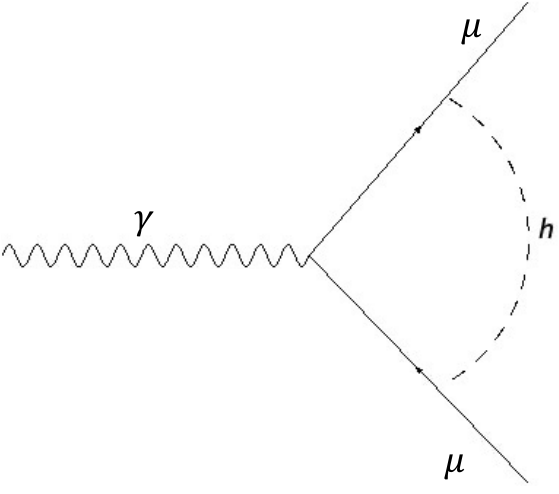}
    \caption{I}
    \label{oneloop}
\end{subfigure}
\hfill
\begin{subfigure}{0.15\textwidth}
    \includegraphics[width=\textwidth]{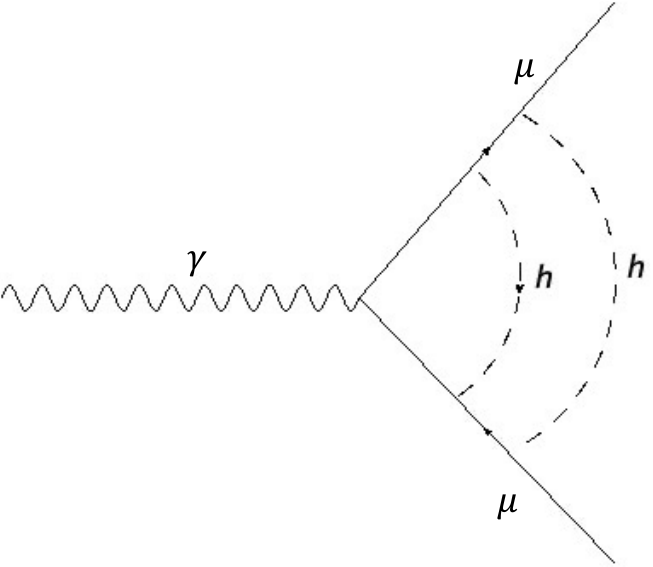}
    \caption{II(a)}
    \label{twoloop1}
\end{subfigure}
\hfill
\begin{subfigure}{0.15\textwidth}
    \includegraphics[width=\textwidth]{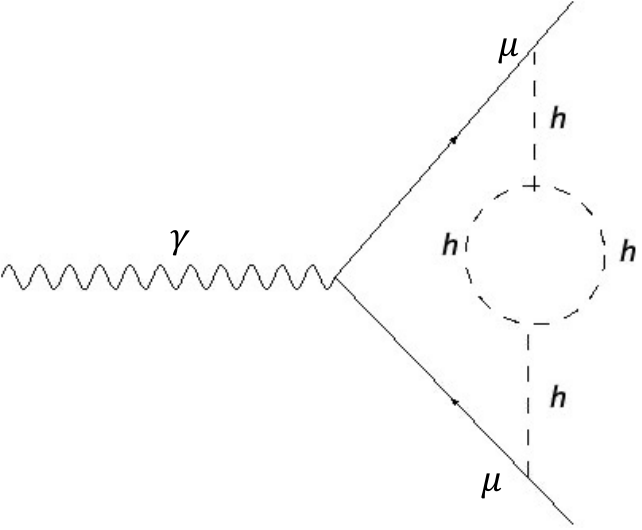}
    \caption{II(b)}
    \label{twoloop2}
\end{subfigure}
\hfill 
     \begin{subfigure}{0.15\textwidth}
    \includegraphics[width=\textwidth]{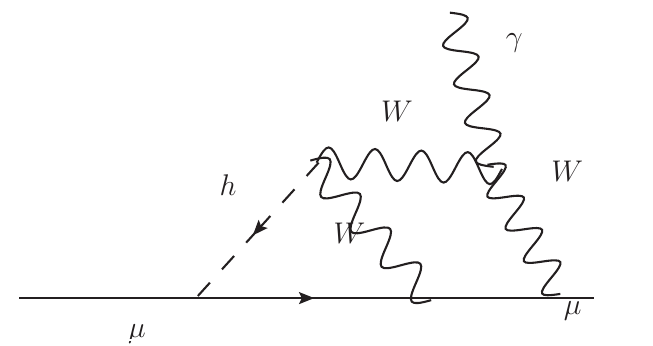}
    \caption{III}
    \label{twoloop3}
\end{subfigure}  
\hfill
 \begin{subfigure}{0.15\textwidth}
    \includegraphics[width=\textwidth]{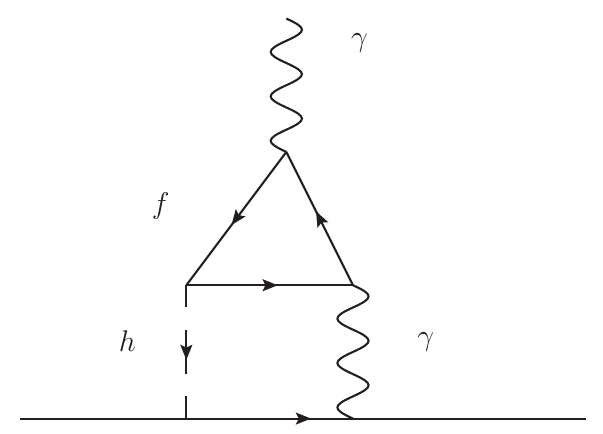}
    \caption{IV}
    \label{twoloop4}
\end{subfigure}  
\hfill
         \begin{subfigure}{0.15\textwidth}
    \includegraphics[width=\textwidth]{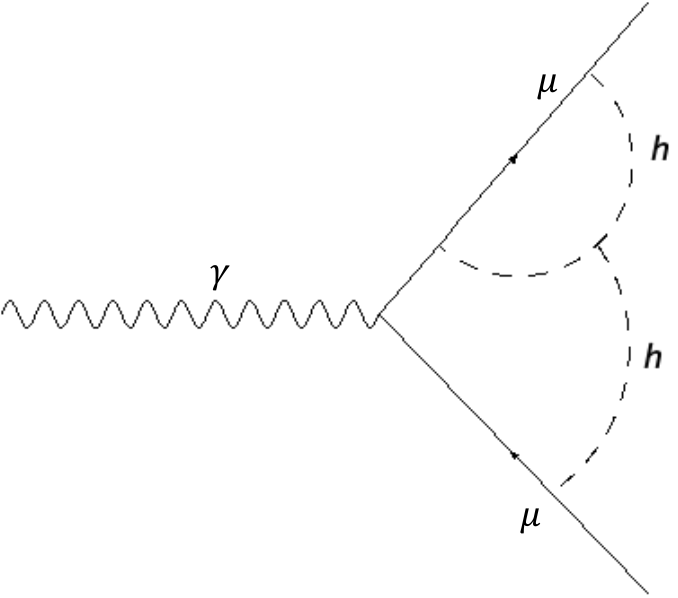}
    \caption{V(a)}
    \label{fig:third}
\end{subfigure}  
\hfill
 \begin{subfigure}{0.15\textwidth}
    \includegraphics[width=\textwidth]{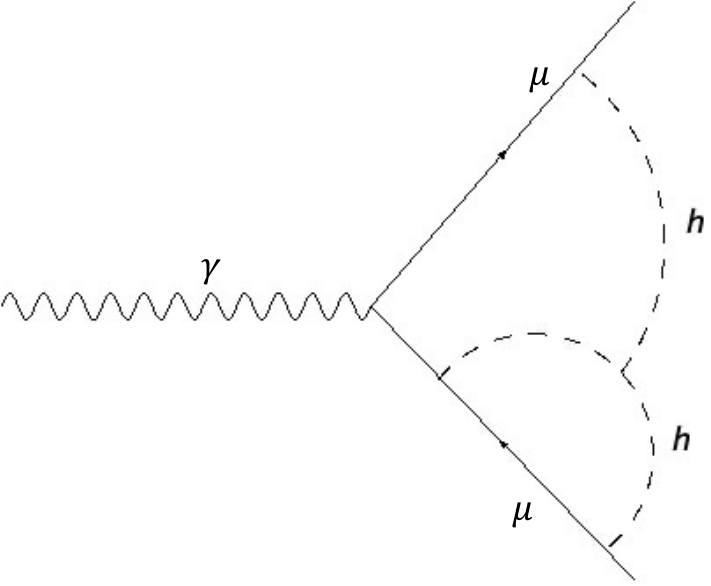}
    \caption{V(b)}
    \label{fig:third}
\end{subfigure}  
\hfill
\begin{subfigure}{0.15\textwidth}
    \includegraphics[width=\textwidth]{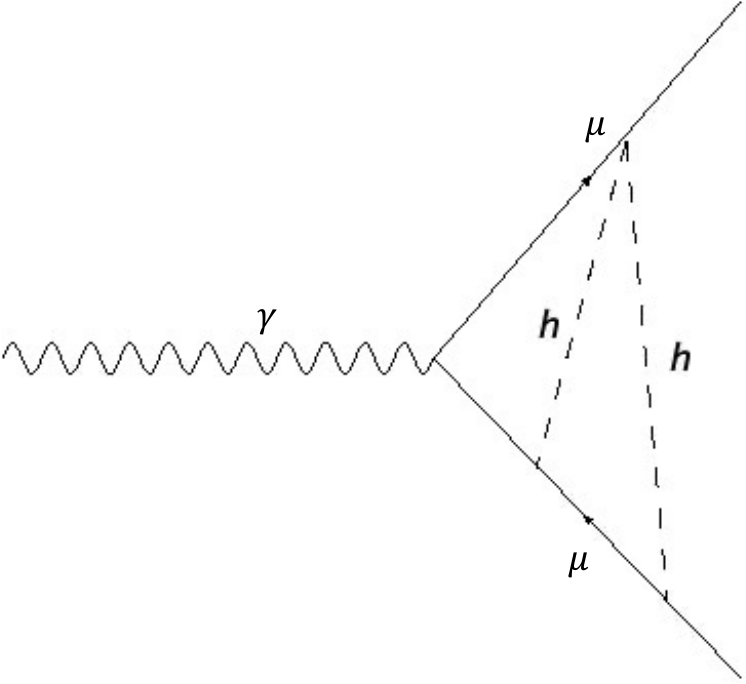}
    \caption{VI(a)}
    \label{fig:third}
\end{subfigure}  
\hfill
\begin{subfigure}{0.15\textwidth}
    \includegraphics[width=\textwidth]{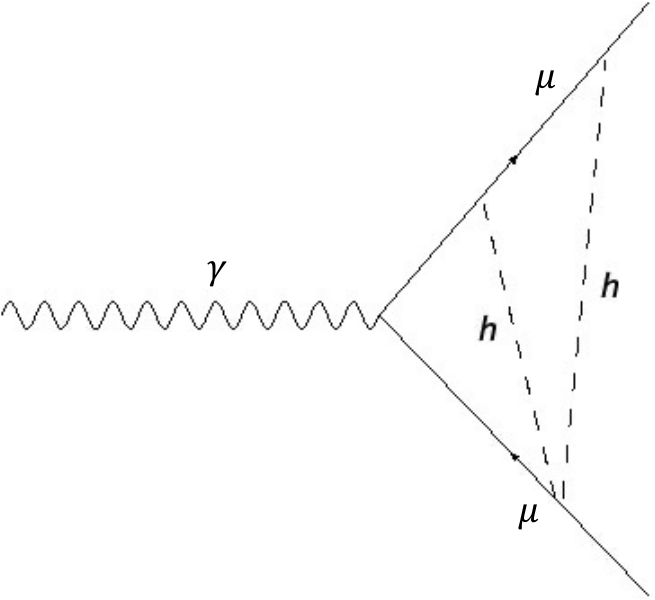}
    \caption{VI(b)}
    \label{fig:third}
\end{subfigure}  
\hfill
\begin{subfigure}{0.15\textwidth}
    \includegraphics[width=\textwidth]{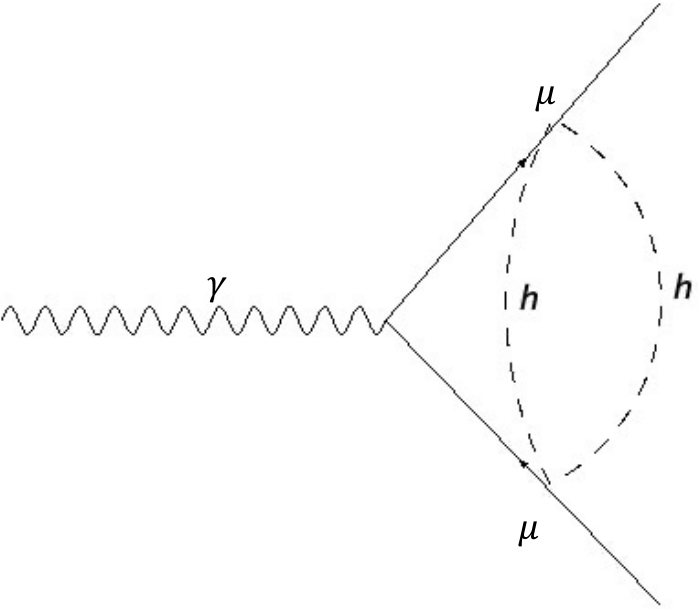}
    \caption{VII}
    \label{fig:third}
\end{subfigure}  
\hfill
\begin{subfigure}{0.15\textwidth}
    \includegraphics[width=\textwidth]{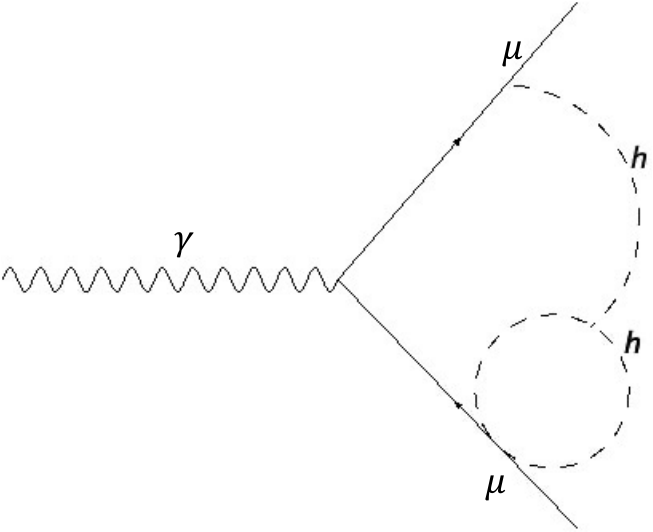}
    \caption{VIII(a)}
    \label{fig:third}
\end{subfigure}  
\hfill
\begin{subfigure}{0.15\textwidth}
    \includegraphics[width=\textwidth]{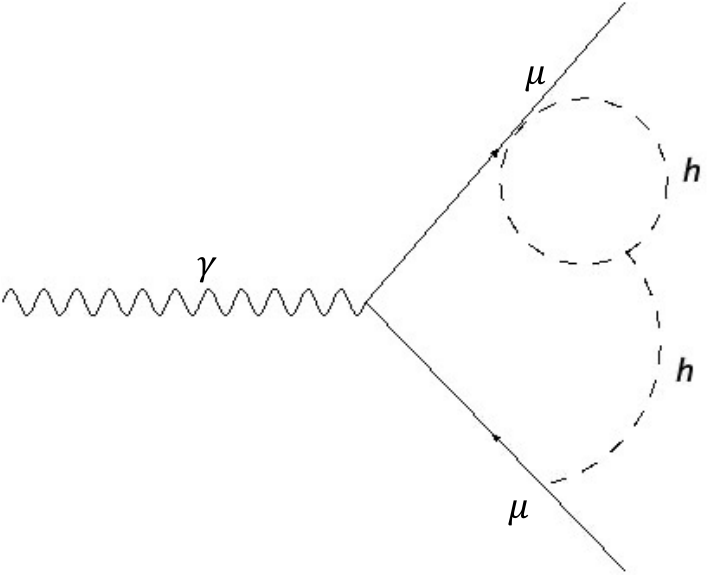}
    \caption{VIII(b)}
    \label{fig:third}
\end{subfigure}   
\caption{\small The diagrams contributing to the $g-2$ of the muon at one and two loops.}
\label{fig1}
\end{figure}

With eq. (\ref{eq:BSM_Lag2}) in hand, we can readily calculate the BSM contributions to $(g-2)_{\mu}$ with $l \rightarrow \mu$, treating the effective couplings as contact terms. In our calculation, we limit ourselves to 2 loops. The Feynman diagrams that contribute to the anomaly at one and two loops are shown in Fig. \ref{fig1}. Diagram I has already been calculated in the SM case \cite{Jackiw:1972jz} (see also \cite{Tucker-Smith:2010wdq, Chen:2015vqy}). Thus, the BSM contribution can be readily calculated simply by substituting $Y_{\mu} \rightarrow \delta_{\mu1} Y_{\mu}$. Given the latest bounds on $\delta_{\mu1}$ \cite{ATLAS:2019nkf}
\begin{equation}\label{eq:Bound}
|\delta_{\mu1}| < 0.53, \hspace{5mm} \text{@ 95\% C.L.},
\end{equation}
it is easy to show that this contribution is at most $O(10^{-14})$ and thus is negligible. Diagram II is loop-suppressed compared to diagram I and therefore is also negligible. Diagrams III and IV (of the Barr-Zee type) have also been calculated before in the SM \cite{Czarnecki:1995sz,Haestier:2006mg, Gribouk:2005ee,Gnendiger:2013pva, Czarnecki:1995wq}. Therefore, we can immediately extend the results to the BSM case with the above rescaling
\begin{align}\label{eq:Contribution1}
\Delta a_{\mu}^{(\text{III})} = \delta_{\mu1} \times a_{\mu,W}^{\text{SM}}, \\
\Delta a_{\mu}^{(\text{IV})} = \delta_{\mu1} \times a_{\mu,f}^{\text{SM}}, 
\end{align}
where $a^{\text{SM}}_{\mu,W} = (-19.97\pm 0.03)\times 10^{-11}$ and $a^{\text{SM}}_{\mu,f} = (-1.5 \pm 0.01) \times 10^{-11}$. Notice here that in the fermion loop in diagram IV, only the contributions from $f = \{t, b, c, \tau\}$ are retained. Also notice in the same diagram that the BSM contribution could include the deviations in the Yukawa couplings of the above fermions, namely $\delta_{t1}, \delta_{b1}, \delta_{c1}$ and $\delta_{\tau1}$, however, given the bounds on these deviations, we checked that their contribution to $\Delta a_{\mu}$ is insignificant, even if we saturate all experimental bounds and arrange for all the corresponding contributions to conspiratorially add up. Therefore, we set $\delta_{t1}, \delta_{b1}, \delta_{c1}, \delta_{\tau1} \rightarrow 0$.

\begin{figure}[!t]
\centerline{\begin{minipage}{0.8\textwidth}
\centering
\centerline{\includegraphics[width=300pt]{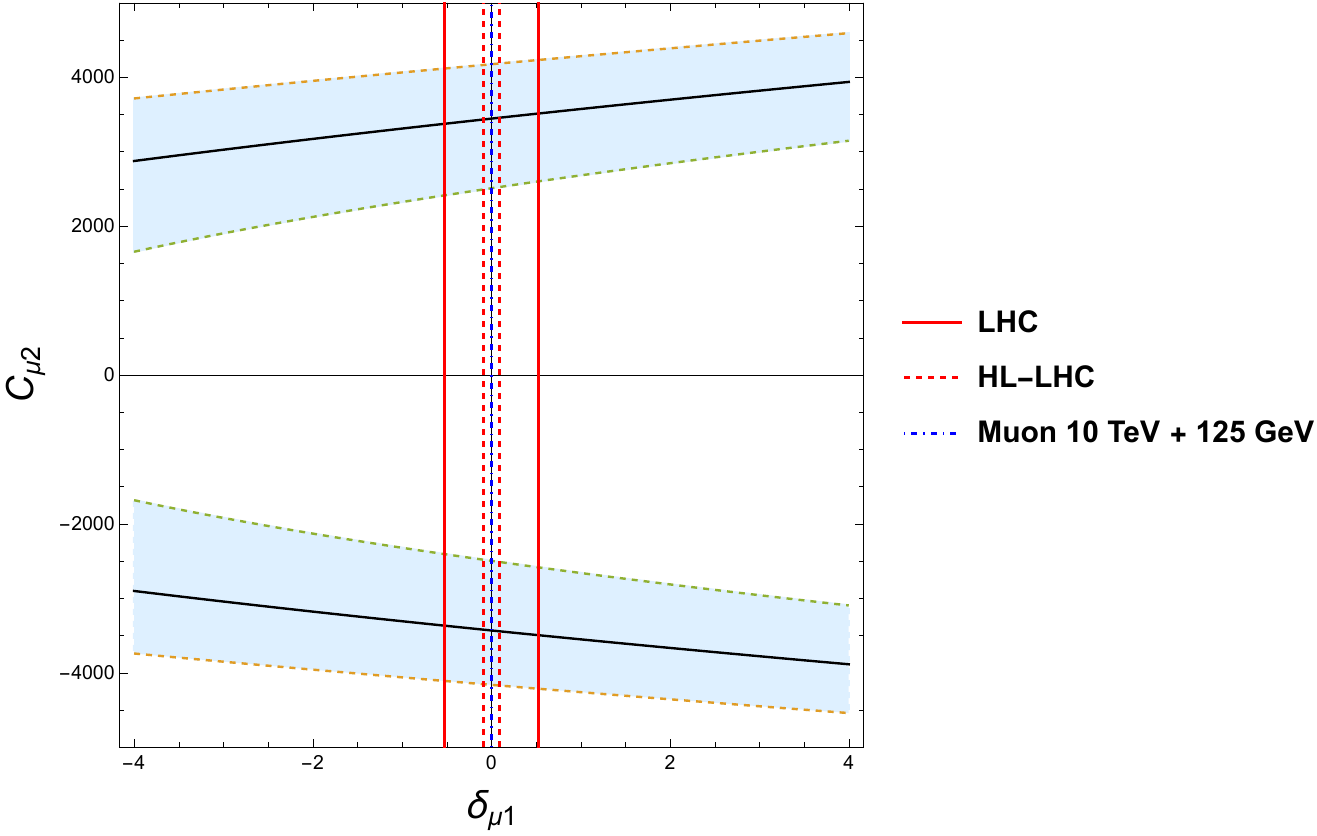}}
\caption{\small The parameter space that corresponds to the $2\sigma$ solution to $(g-2)_{\mu}$, together with the LHC constraints (solid red), the HL-LHC projections (dotted red), and the muon collider projections (dotted blue) on $\delta_{\mu1}$, superimposed. $C_{\mu2}$ is unconstrained.}
\label{fig2}
\end{minipage}}
\end{figure}

Diagram V only depends $\delta_{\mu1}$, and explicit calculation of its contribution shows that it is of $O(10^{-18})$ when the bound in eq. \ref{eq:Bound} is saturated. On the other hand, the contribution from diagram VI depends on both $\delta_{\mu1}$ and $C_{\mu2}$ and therefore could be significant. Nonetheless, we checked that compared to the other diagrams, the contribution is suppressed and does not exceed $O(10^{-13})$ at best, and thus can safely be neglected. 

This leaves us with diagrams VII and VIII. Both diagrams are somewhat complicated, however, if we use a UV-cutoff scheme such that $\Lambda \gg M_{h}$, then one can show that they can be approximated by
\begin{eqnarray}\label{eq:Contribution2}
\Delta a_{\mu}^{(\text{VII})} & \simeq & \frac{m_{\mu}^{4}}{128\pi^{4}v^{4}}C_{\mu2}^{2} \log^{2}{\Big(\frac{\Lambda^{2}}{M_{h}^{2}}} \Big), \\
\Delta a_{\mu}^{(\text{VIII})} & \simeq & -\frac{3m_{\mu}^{4}}{64\pi^{4}v^{4}}(1+\delta_{\mu1}) C_{\mu2} \log{\Big(\frac{\Lambda^{2}}{M_{h}^{2}}} \Big)\log{\Big(\frac{M_{h}^{2}}{m_{\mu}^{2}}} \Big).
\end{eqnarray}

Here we note in passing that we are only using a UV cutoff scheme for convenience, and using dimensional regularization yields similar results if the renormalization scale is defined appropriately. The results are insensitive to the choice of the renormalization scheme or the UV cutoff/ renormalization scale, as the divergence is only logarithmic. Putting all pieces together, the total BSM contribution to $\Delta a_{\mu}$ is thus given by
\begin{equation}\label{eq:TotalBSM}
\Delta a_{\mu}^{\text{BSM}} \simeq \Delta a_{\mu}^{(\text{III})} + \Delta a_{\mu}^{(\text{IV})} + \Delta a_{\mu}^{(\text{VII})} + \Delta a_{\mu}^{(\text{VIII})}.
\end{equation}

We have 3 parameters, namely $\delta_{\mu1}$, $C_{\mu2}$ and $\Lambda$. In the rest of this paper, we fix $\Lambda = 10$ TeV. Now, we can use eq. (\ref{eq:TotalBSM}) in order to find the region in the parameter space that corresponds to solving the anomaly in eq. \ref{eq:DeltaaExp}. Figure \ref{fig2} shows the viable parameter space that corresponds to the $2\sigma$ solution to $\Delta a_{\mu}$. The plot also shows the experimental constraints given in eq. \ref{eq:Bound} in solid red. The plot shows that the Higgs sector can still accommodate a solution to the $(g-2)_{\mu}$ anomaly. This is mainly because the BSM Wilson coefficient $C_{\mu2}$ is essentially unconstrained experimentally, and we can see that a solution to the $g-2$ anomaly is still possible even if $\delta_{\mu1} = 0$. On the other hand, $\delta_{\mu1}$ cannot on its own solve the anomaly if $C_{\mu2} \rightarrow 0$. In fact, we can see that $\delta_{\mu1}$ only mildly affects $\Delta a_{\mu}$, which is hardly surprising given the smallness of the muon Yukawa coupling and the bound in eq. \ref{eq:Bound}.

 The size of $|C_{\mu2}|$ required to solve the anomaly is $\sim (2-4) \times 10^{3}$, which might seem large, however, we should keep in mind that this is merely an artifact of our parameterization when we divided $C_{\mu2}$ by $v$ to keep it dimensionless. We should note that with proper power counting (such as in the SMEFT), $C_{\mu2}$ is divided by a larger UV scale that is typically $O(\text{TeV})$ which would yield lower values of $|C_{\mu2}|$. We shall show later on that perturbativity is maintained for this range of $|C_{\mu2}|$ up to safe levels. In addition, we shall discuss the fine-tuning associated with it in the next section.

\begin{table}[!t]
\centering
    \begin{tabular}{|c|c|c|c|c|c|c|c|c|}
    \hline
& \text{HL-LHC} & \text{HE-LHC} & \text{ILC} & \text{CLIC} & \text{CEPC} & \text{FCC-ee} & \text{FCC} & \text{Muon (10 TeV)}\\
&  & \text{S2} \hspace{1mm} \text{S2}$^{\prime}$ & \text{250} \hspace{1mm} \text{500}  \hspace{1mm} \text{1000} & \text{380}  \hspace{1mm} \text{1500}  \hspace{1mm} \text{3000} & & \text{240}  \hspace{1mm} \text{365} & \text{(ee/eh/hh)} & \hspace{5mm} \text{+125 GeV}\\
    \hline
$\delta_{\mu1} [\%]$ & $9.2$ & $5$ \hspace{1mm} $3.4$ & $30$ \hspace{1mm} $18.8$ \hspace{1mm} $12.4$ & $640$ \hspace{2mm} $26$ \hspace{2mm} $11.6$ & $17.8$ & $20$ \hspace{1mm} $17.8$ & $0.82$ & $3.6$ \hspace{5mm} $0.19$ \\
    \hline
    \end{tabular}
    \caption{\small Projected $2\sigma$ bounds on $\delta_{\mu1}$ in the various future experiments taken from the \text{Higgs@FutureColliders} study \cite{deBlas:2019rxi} which is summarized in \cite{Abu-Ajamieh:2022dtm} (see also \cite{Dawson:2022zbb}). The most stringent constraints come from the combined analyses in the muon collider with COM energies of $10$ TeV and $125$ GeV. This experiment is essentially capable of exploring the entire $\delta_{\mu1}$ dependence of the parameter space.}
    \label{Table2}
\end{table}

The current LHC bound in eq. \ref{eq:Bound} is not very stringent, however, there are many proposed future experiments that are expected to probe the viable parameter space almost fully. These are summarized in Table \ref{Table2}. We show the most relevant ones in Figure \ref{fig2}, which are the HL-LHC projections and the projections from the proposed muon collider with COM runs of $10 \hspace{1mm} \text{TeV} + 125 \hspace{1mm} \text{GeV}$. We can see from the plot that the proposed muon collider can essentially probe the entire range of $\delta_{\mu1}$. However, even in this scenario, the Higgs sector remains viable as a potential solution to the $g-2$ anomaly, because $C_{\mu2}$ remains unconstrained. To the best of our knowledge, there aren't any proposed measurements of $C_{\mu2}$ either currently in the LHC or in the future. We think that current and future experiments should probe this effective coupling both because it can potentially provide a solution to the $g-2$ anomaly, even when $\delta_{\mu1} \rightarrow 0$; and because of its potential enhancement of the di-Higgs production as we will show below.

\section{Phenomenological implications}\label{Sec:Pheno}
In this section, we will discuss the phenomenological implications of the viable parameter space, including the implications of nonzero $\delta_{\mu1}$ and $C_{\mu2}$ on unitarity; corrections to the muon mass and the corresponding fine-tuning, the impact on other couplings; and the enhancement to the di-Higgs production.

\subsection{Unitarity-Violation and the Scale of New Physics}\label{sec:Unitarity}
As explained in detail in \cite{Chang:2019vez, Abu-Ajamieh:2020yqi, Abu-Ajamieh:2021egq,Abu-Ajamieh:2022ppp, Abu-Ajamieh:2021vnh}, any non-vanishing deviations from the SM predictions will eventually lead to a breakdown in unitarity at some high energy scale. This is because the SM is the unique UV-complete theory with the observed particle content, and any deviation from the SM predictions will lead to UV-incompleteness that manifests itself as processes with energy-growing amplitudes that eventually lead to a breakdown in unitarity at some high energy scale, which signals the onset of NP. 

This implies that each point in the parameter space that corresponds to $\Delta a_{\mu}$ will point to an energy scale where unitarity is violated, and thus to the scale at which NP should come into play. It is fairly easy to adapt the results of the top sector in \cite{Abu-Ajamieh:2022ppp} to the muon sector. There is only one model-independent process, namely
\begin{equation}\label{eq:process1}
    \mu \bar{\mu} \rightarrow hh: E_{\text{max}} = \frac{8\sqrt{2}\pi v^{2}}{m_{\mu} C_{\mu2}}.
\end{equation}

\begin{figure}[!t]
\centering
\begin{minipage}{0.45\textwidth}
  \centering
  \includegraphics[width=\linewidth]{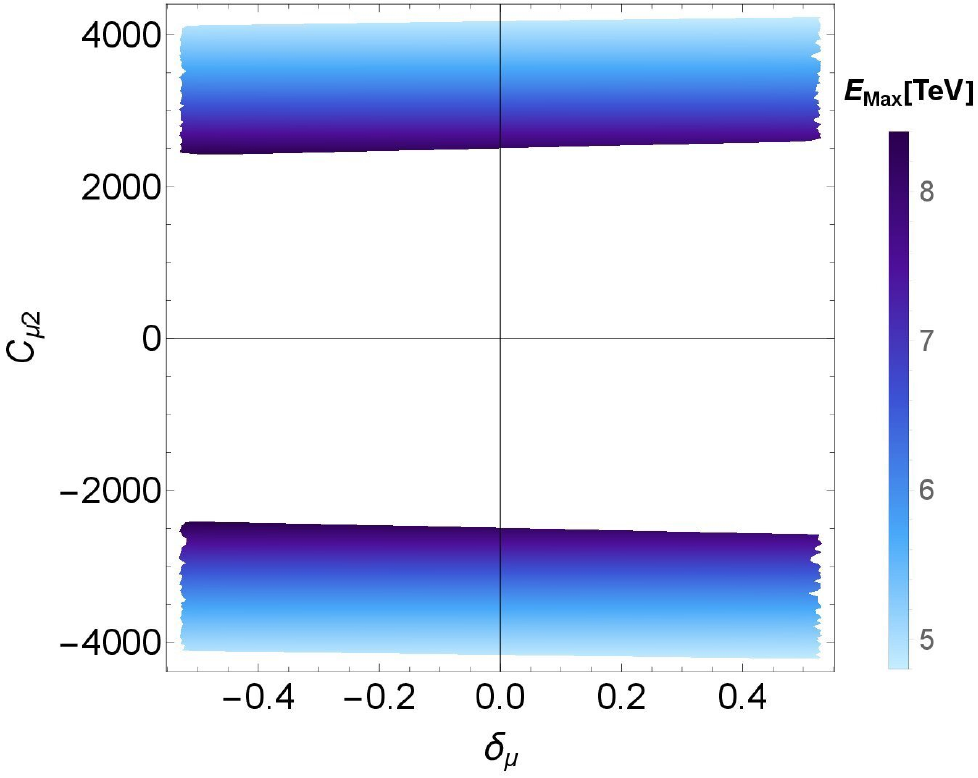}
\end{minipage}%
\begin{minipage}{0.45\textwidth}
  \centering
  \includegraphics[width=
  \linewidth]{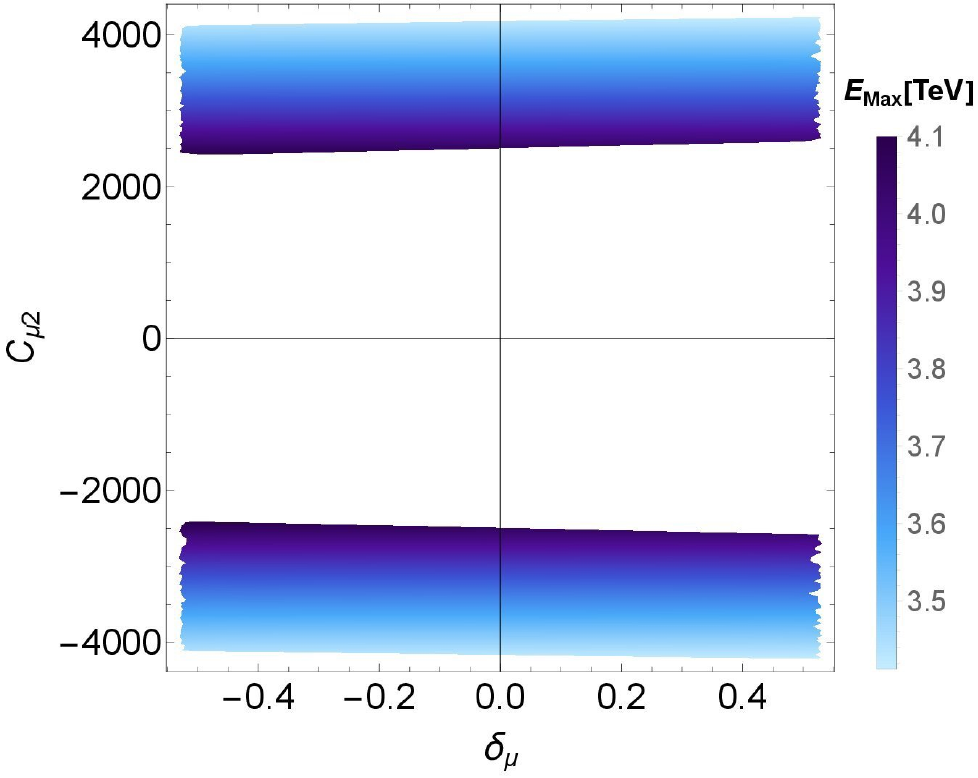}
\end{minipage}
\caption{\small The model-independent unitarity-violating scale that corresponds to the parameter space that can solve the $g-2$ anomaly when the Higgs couplings to the $W$ and $Z$ are allowed to deviate from the SM predictions (left) and when they conform to the SM predictions (right). In the former case, the scale of NP is $\sim 4.8 - 8.4-$ TeV, whereas in the latter case, it could be as low as $\sim 3.4 - 4$ TeV.}
\label{fig3}
\end{figure}

This process is model-independent because it only depends on one parameter, i.e. $C_{\mu2}$, and thus the scale of NP is independent of any other assumptions regarding the other potential deviations in the SM. We show the scale of NP that corresponds to the parameter space obtained from eq. \ref{eq:process1} on the left side of Figure \ref{fig3}. We can see from the plot that the scale of NP is $\sim 4.8 - 8.4$ TeV, which could be probed in the LHC, although if the scale of NP lies at the higher end, this might be difficult. However, this is well within the range of the future $100$-TeV collider, and possibly the proposed muon collider.

Better bounds, however, can be obtained if the deviations in the couplings of the Higgs to the massive gauge bosons are assumed to vanish. As shown in \cite{Abu-Ajamieh:2022ppp}, there are other process that violate unitarity, however, the amplitudes of these processes also receive contamination from the deviation in $hhV$ and $hhVV$ couplings, with $V$ being the $W$ or the $Z$. Therefore, the scale of NP will also depend on the sizes of these deviations. However, if these deviations are assumed to vanish, then these processes will become functions of $\delta_{\mu1}$ and $C_{\mu2}$ only, which enables us to obtain a lower scale of NP. Up to 6-body scattering, these processes include\footnote{We refer the interested reader to \cite{Abu-Ajamieh:2022ppp} for a complete discussion of how these bounds are obtained from unitarity.}
\begin{eqnarray}
\mu \bar{\mu} \rightarrow W_{L}^{+}W_{L}^{-} & : & E_{\text{max}} \leq \frac{8\pi v^{2}}{m_{\mu}\delta_{\mu1}}, \label{eq:processesa}\\
\mu \bar{\nu} \rightarrow W_{L}^{-}h & : & E_{\text{max}} \leq \frac{8\pi v^{2}}{\sqrt{2} m_{\mu}\delta_{\mu1}},\label{eq:processesb}\\
\mu \bar{\nu} \rightarrow W_{L}^{-}W_{L}^{+}W_{L}^{-} & : & E_{\text{max}} \leq \Big[ \frac{32\pi^{2} v^{3}}{\sqrt{2}m_{\mu}\delta_{\mu1}}\Big]^{\frac{1}{2}},\label{eq:processesc}\\
\mu \bar{\mu} \rightarrow W_{L}^{+}W_{L}^{-}h, \hspace{1mm} \mu \bar{\nu} \rightarrow W_{L}^{-} h^{2} & : & E_{\text{max}} \leq \Big[ \frac{64 \pi^{2} v^{3}}{\sqrt{2}m_{\mu}|C_{\mu2}-\delta_{\mu1}|}\Big]^{\frac{1}{2}}\label{eq:processesd},\\
\mu \bar{\nu}W_{L}^{+} \rightarrow W_{L}^{+}W_{L}^{-}h & : & E_{\text{max}} \leq \Big[\frac{128 \sqrt{3} \pi^{3} v^{4}}{\sqrt{2}m_{\mu}|C_{\mu2}-3\delta_{\mu1}|} \Big]^{\frac{1}{3}}.\label{eq:processese}
\end{eqnarray}
where $W_{L}^{\pm}$ are the longitudinal modes of the $W$ boson. We plot the lowest unitarity-violating scale from eqs. (\ref{eq:processesa}-\ref{eq:processese}) on the right-hand side of Figure \ref{fig3}. We can see now that the scale of NP drops to $\sim 3.4 - 4$ TeV, which is indeed within the reach of the LHC. This suggests that the HL-LHC could shed more light on the potential BSM physics in the Higgs sector and its potential solution of the $g-2$ anomaly.

\subsection{Corrections to the Muon Mass and Fine-tuning}
Nonzero $\delta_{\mu1}$ and/or $C_{\mu 2}$ will yield corrections to the muon mass. Given the bound in eq. \ref{eq:Bound}, it is easy to see that $\delta_{\mu 1}$ will at most lead to a mass correction $\delta m_{\mu} \sim 2\delta_{\mu 1} m_{\mu}$, and thus to fine-tuning $\sim O(1)$. On the other hand, the required size of $C_{\mu 2}$ to solve the anomaly is larger, and thus the corresponding mass correction could be significantly larger.

The correction to the muon mass due to $C_{\mu2}$ arises when the Higgs loop is closed. Thus, calculating the mass correction can easily be done by integrating out that loop. The corresponding fine-tuning can be estimated as
\begin{equation}\label{eq:fine-tuning}
    \frac{\delta m_{\mu}}{m_{\mu}} \sim C_{\mu2}\frac{M^{2}_{h}}{32\pi^{2}v^{2}},
\end{equation}

Given that the size of $C_{\mu2}$ required to solve the anomaly is $\sim (2-4)\times 10^{3}$, is is easy to see that although the fine-tuning is larger in this case, it is nonetheless still of $O(1)$, and thus remains acceptable.


\subsection{Modification to Muon's Coupling to the $W$ and $Z$}\label{sec:Mod_WZ}
The Lagrangian in eq. (\ref{eq:BSM_Lag1}) might suggest that the couplings of the muon to other SM particles might be significantly modified. In particular, when eq. (\ref{eq:BSM_Lag1}) is expressed in a general gauge where the Goldstone bosons are manifest, one would expect from the equivalence theorem, that the coupling of $\mu$ to $W$ and $Z$ will be modified compared to their SM predictions. Nonetheless, here we show that this is not the case, as we are defining the field $X$ to be VEV-subtracted as can be seen in eq. (\ref{eq:Xfield}). To show this explicitly, we write the Higgs doublet in a general gauge
\begin{equation}\label{eq:Higgs_General}
H = \frac{1}{\sqrt{2}} \begin{pmatrix} 
		G_{1} + i G_{2}\\
		v + h + i G_{3}
		\end{pmatrix}
\end{equation}
then we insert $H$ in eqs. (\ref{eq:Xfield}) and (\ref{eq:Projector}) and expand in powers of $\vec{G}$, we find
\begin{eqnarray}\label{eq:G_expansion}
X & = & \sqrt{(v+h)^{2}+G_{1}^{2}+G_{2}^{2}+G_{3}^{2}} - v \simeq h + O(\vec{G)},\\
\Tilde{\hat{H}} & \simeq & \begin{pmatrix} 
\frac{-G_{1}+ i G_{2}}{v+h}\\
-1 - \frac{i G_{3}}{v+h} 
\end{pmatrix} + O(\vec{G}^{2}),
\end{eqnarray}
and the eq. (\ref{eq:BSM_Lag1}) becomes
\begin{equation}\label{eq:Lag_General_Gauge}
\delta \mathcal{L}_{\text{eff}} \simeq \frac{Y_{\mu}\delta_{\mu 1}}{\sqrt{2}}\Big( \frac{h}{v+h}\Big)\Big[\sqrt{2} \overline{\nu}_{\mu L}\mu_{R}W_{L}^{+} + \sqrt{2} \overline{\mu}_{L}\nu_{\mu R}W_{L}^{-}  + i\overline{\mu}\mu Z_{L} \Big].
\end{equation}

We can expand eq. (\ref{eq:Lag_General_Gauge}) in powers of $h/v$, where it is easy to see that at leading order, there is no correction to either $\mu\mu Z$ or $\mu\nu W$. On the other hand, we could have the effective couplings $\mu\mu Z h$ and $\mu\nu W h$. Nonetheless, it is easy to see that even when $\delta_{\mu1}$ saturates the bound in es. (\ref{eq:Bound}), these couplings are $\sim O(10^{-6})$ at most, and thus can easily evade any experimental constraints.

\subsection{Di-Higgs Production Enhancement}\label{sec:diHiggs}
Nonzero $\delta_{\mu1}$ and/or $C_{\mu2}$ could have interesting phenomenological consequences, such as deviation in the branching fraction of the Higgs decay to $\mu^{+}\mu^{-}$, and deviations in the Higgs signal strength from the SM predictions. However, perhaps the most interesting experimental signature for a nonzero $\delta_{\mu1}$ and/or $C_{\mu2}$ is their enhancement of the di-Higgs production. The di-Higgs production provides an excellent probe for NP through both investigating the Higgs couplings compared to the SM, and searching for BSM particles through resonant enhancement, such as additional heavy Higgses, KK gravitons etc. Thus, we confine our discussion to the potential enhancement of the di-Higgs production. 

\begin{figure}[!t]
\centerline{\begin{minipage}{0.6\textwidth}
\centering
\centerline{\includegraphics[width=250pt]{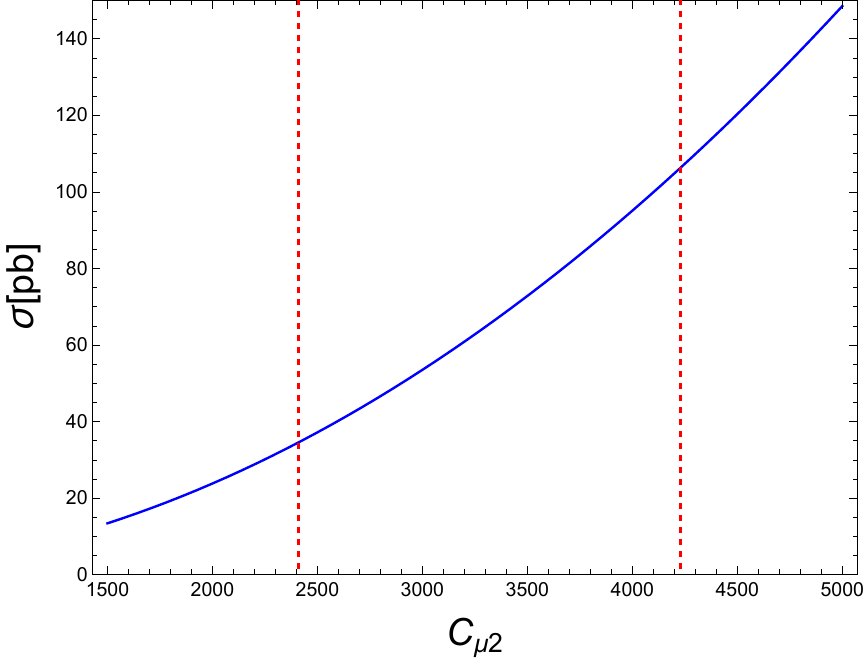}}
\caption{\small The cross-section of the di-Higgs production in the proposed $10$-TeV muon collider as a function of $C_{\mu2}$. The red dashed lines bound the region of the parameter space that corresponds to solving the $g-2$ anomaly.}
\label{fig4}
\end{minipage}}
\end{figure}

The strongest enhancement of the di-Higgs production through nonzero $\delta_{\mu1}$ and/or $C_{\mu2}$ arises in muon colliders. In a muon collider, the di-Higgs production proceeds at tree-level and is therefore expected to be sizable. Specifically, $\mu \bar{\mu} \rightarrow hh$ proceeds through the $s, t$ and $u$ channels, in addition to the contact term. The $s, t$ and $u$ channels are proportional to $\delta_{\mu1}$ alone and are negligible. On the other hand the contact term depends on $C_{\mu2}$ and can be sizable. Its cross-section is given by
\begin{equation}
\sigma^{\text{BSM}}(\mu\bar{\mu} \rightarrow hh) \simeq \frac{C_{\mu2}^{2}}{64\pi v^{4}} m_{\mu}^{2}\sqrt{1-\frac{4M_{h}^{2}}{s}}.
\end{equation}

In Figure \ref{fig4}, we show the cross-section of the di-Higgs production in the proposed $10$-TeV muon collider. The plot shows that for the region of the parameter space that corresponds to the solution to the $g-2$ anomaly, the di-Higgs production cross-section is $\sim 30-105$ pb, which is quite sizable. This could be the smoking gun for the viability of the Higgs solution to the $g-2$ anomaly, and it indeed represents another motivation for building the muon collider.

\section{A Possible UV Model}\label{sec:UVmodels}
So far, we have followed a completely bottom-up model-independent approach without reference to any UV completions. Similar work has been discussed in the literature using the SMEFT, see for instance \cite{Dermisek:2022aec}, where it was argued that any new interaction that results in a chirally-enhanced contribution to $(g-2)_{\mu}$, necessarily modifies the decay rate of the Higgs to $\overline{\mu}\mu$, or leads to an electric dipole moment for the muon.

In our paper, we have seen that a proper solution would require sizable contributions to the $\mu\mu h^{2}$, while at the same time keep the contribution to $\mu\mu h$ small to evade the LHC constraints in eq. (\ref{eq:Bound}). Building such a model is challenging as it would require conspiratorial cancellations to achieve such a requirement. Here we present a model than can achieve this using vectorlike leptons. The use of vectorlike leptons to solve the $g-2$ anomaly is not new, and has been studied extensively in the literature (see for instance \cite{Dermisek:2021mhi,Kannike:2011ng, Dermisek:2013gta, Poh:2017tfo, Dermisek:2021ajd, Arkani-Hamed:2021xlp}).

We adopt a model similar to the one presented in \cite{Dermisek:2021ajd}. The SM is extended by two vectorlike $SU(2)$ doublets $L_{L,R}$ and two vectorlike $S(2)$ singlets $E_{L,R}$. $L_{L}$ and $E_{R}$ have the same quantum numbers as the SM leptons. In addition, we also introduce a heavy Higgs-like scalar doublet $\Phi$ that develops a VEV $v_{\phi}$
\begin{equation}\label{eq:Phi_doublet}
\Phi = \begin{pmatrix}
0\\
v_{\phi} + \phi
\end{pmatrix}.
\end{equation}

\begin{figure}[!t]
\centerline{\begin{minipage}{0.6\textwidth}
\centering
\centerline{\includegraphics[width=300pt]{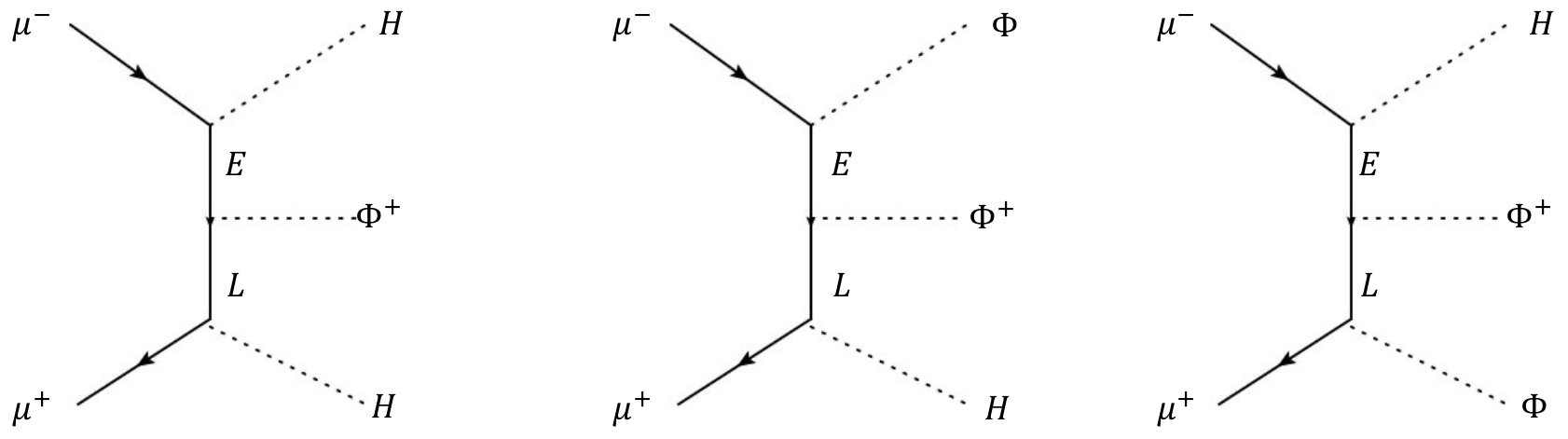}}
\caption{\small The relevant UV processes. After integrating out the fields $L$ and $E$, we obtain the effective Lagrangian in eq. (\ref{eq:UV_effective}).}
\label{fig5}
\end{minipage}}
\end{figure}

The most general Lagrangian can be written as
\begin{equation}\label{eq:UV_Lagrangian}
\begin{split}
\mathcal{L}_{\text{UV}} = & -M_{L}\overline{L}_{L}L_{R} -M_{E}\overline{E}_{L}E_{R} - y_{\mu}\overline{l}_{L}\mu_{R}H - Y_{\mu} \overline{l}_{L}\mu_{R} \Phi -\lambda_{E}\overline{l}_{L}E_{R}H -\lambda_{L}\overline{L}_{L}\mu_{R}H \\
& - \lambda\overline{L}_{L}E_{R}H - \overline{\lambda}H^{\dagger}\overline{E}_{L}L_{R} -\kappa_{E}\overline{l}_{L}E_{R}\Phi - \kappa_{L}\overline{L}_{L}\mu_{R}\Phi - \kappa \overline{L}_{L}E_{R}\Phi - \overline{\kappa}\Phi^{\dagger}\overline{E}_{L}L_{R} + \text{h.c.},
\end{split}
\end{equation}
where we have neglected the interaction of $\Phi$ with $H$ as they are irrelevant for our purposes. However, we assume that the parameters of the scalar potential are such that they are consistent with experiment. In our model, we set $\overline{\lambda} \rightarrow 0$. Assuming that $M_{L}, M_{E} \gg M_{\text{EW}}$, we can integrate out $L$ and $E$ (see Figure \ref{fig5}) to arrive at the effective Lagrangian
\begin{equation}\label{eq:UV_effective}
	\mathcal{L}_{\text{EFT}} \supset - y_{\mu}\overline{l}_{L}\mu_{R}H - Y_{\mu} \overline{l}_{L}\mu_{R}\Phi -\frac{\lambda_{L}\overline{\kappa}\lambda_{E}}{M_{L}M_{E}}\overline{l}_{L}\mu_{R}H\Phi^{\dagger}H -\overline{\kappa} \Bigg( \frac{\kappa_{E}\lambda_{L}+ \lambda_{E}\kappa_{L}}{M_{L}M_{E}} \Bigg)\overline{l}_{L}\mu_{R}\Phi\Phi^{\dagger}H + h.c.,
\end{equation}

Defining $\overline{\kappa}/M_{L}M_{E} \equiv 1/\overline{M}$ and matching the Lagrangian in eq. (\ref{eq:UV_effective}) with the effective Lagrangian in eq. (\ref{eq:BSM_Lag2}), we obtain the following matching conditions:
\begin{eqnarray}\label{eq:Matching_conditions}
\delta m_{\mu} & = & v_{\phi}Y_{\mu} + \frac{vv_{\phi}}{2\overline{M}^{2}}\Big(v \lambda_{L}\lambda_{E} + \sqrt{2}v_{\phi}(\kappa_{E}\lambda_{L}+\lambda_{E}\kappa_{L})\Big),\\
\delta_{\mu1} & = & \frac{vv_{\phi}}{\sqrt{2}\hspace{1mm}\overline{M}^{2}m_{\mu}}\Big(\sqrt{2}v \lambda_{L}\lambda_{E} + v_{\phi}(\kappa_{E}\lambda_{L}+\lambda_{E}\kappa_{L})\Big),\\
C_{\mu 2} & = & \frac{v^{2}v_{\phi}\lambda_{L}\lambda_{E}}{\overline{M}^{2}m_{\mu}}.
\end{eqnarray}

In order to be consistent with experimental measurements, we require that $\delta m_{\mu} = 0$, which leads to
\begin{equation}\label{eq:C2_matched}
	C_{\mu2} = 2\Big( \frac{v_{\phi}Y_{\mu}}{m_{\mu}} + \delta_{\mu1} \Big).
\end{equation}
and we can see that we can easily generate the required hierarchy in the scales of the $\mu\mu h$ and $\mu\mu h^{2}$ couplings with $v_{\phi} \sim$ TeV. For instance, setting $ \delta_{\mu1} = 0$, $Y_{\mu} = 0.1$, we can generate $C_{\mu 2} \sim (2-4) \times 10^{3}$ by selecting $v_{\phi} \sim 1-2$ TeV. 

We can see from the proposed UV completion, that maintaining small corrections at LO while allowing for sizable corrections at NLO, would require some conspiratorial cancellations. Nonetheless, one can envisage that such cancellations might arise from a hidden symmetry that protects the LO from receiving large corrections. We will not pursue this issue any further.

\section{Conclusions and outlook}\label{Sec:Conc}
In this paper, we analyzed the viability of a BSM Higgs solution to the $g-2$ anomaly using a bottom-up EFT approach. In the SM, the Higgs contribution to the muon's magnetic dipole moment is negligible. However, given that the current level of experimental limits on the Higgs sector still leaves ample room for NP, it is possible for this contribution to be significantly enhanced compared to the SM.

We saw that while the deviation in the Higgs Yukawa coupling to the muon is not enough to account for the anomaly, the effective coupling $\bar{\mu}\mu hh$, which is unconstrained experimentally, can account for the entire discrepancy. We surveyed the proposed projections on the parameter space and found that the proposed muon collider could provide the best sensitivity for exploring the parameter space. We also showed that unitarity can be used in order to estimate the scale of NP that corresponds to non-vanishing deviations from the SM, i.e. $\delta_{\mu1}$ and $C_{\mu2}$. We found that the model-independent scale of NP that corresponds to the region of the parameter space that can solve the anomaly to be $\sim 5 - 8$ TeV. This scale is lowered to $\sim 3.4 - 4$ TeV if the Higgs couplings to massive gauge bosons are assumed to be equal to their SM predictions. We also showed that the corrections to the muon mass that correspond to the size of $\delta_{\mu1}$ and $C_{\mu2}$ required to solve the anomaly, correspond to a fine-tuning level of $O(1)$ only.

We also discussed the possible enhancement of the di-Higgs production cross-section, and found that it could be sizable in the proposed $10$ TeV muon collider as a result of the contact term interaction through $C_{\mu2}$. A key takeaway of this study is that current and future experiments should try to measure and set limits on $C_{\mu2}$ due to its potential significance in solving the $g-2$ anomaly. Current searches focus of the $hh\bar{t}t$ coupling $C_{t2}$, which if non-vanishing, is expected to be the largest of the effective couplings of the type $hh\bar{f}f$. Nonetheless, we believe that $C_{\mu2}$ should be given similar interest. 

In this paper, we confined our analysis to flavor-conserving operators and avoided Flavor Violation (FV). FV Yukawa couplings $Y_{ij} \neq \delta_{ij}\sqrt{2}m_{i}/v$ have been discussed extensively in the literature, see for instance \cite{Harnik:2012pb} and the references therein, however, to the best of our knowledge, FV effective operators such as $C_{ij}$ have never been properly investigated before, although they were briefly discussed in \cite{Babu:1999me, Giudice:2008uua, Goudelis:2011un}. We will study these types of FV coupling in a future work.
\vspace{0pt}
\section*{Acknowledgments}
FA would like to that Stephen Martin for the valuable correspondence. The work of FA is supported by the C.V. Raman fellowship from CHEP at IISc. S.K.V. thanks SERB Grant CRG/2021/007170 "Tiny Effects from Heavy New Physics" from Department of Science and Technology, Government of India.

\end{document}